\documentclass[sigconf, nonacm]{acmart}

\usepackage{amsmath}
\usepackage{newtxtext,newtxmath}
\usepackage[ruled,linesnumbered]{algorithm2e}
\usepackage{svg}
\usepackage{color, colortbl}
\usepackage{xcolor}
\usepackage{caption}
\usepackage{subcaption}
\usepackage{balance}
\usepackage{multicol}
\usepackage{multirow, makecell, caption}
\usepackage{graphicx}

\definecolor{Gray}{gray}{0.9}

\DeclareMathOperator{\srzero}{\mathbf{0}}
\DeclareMathOperator{\srtimes}{\otimes}
\DeclareMathOperator{\srplus}{\oplus}

\DeclareMathOperator{\mplus}{\srplus}
\DeclareMathOperator{\mtimes}{\srtimes}
\DeclareMathOperator{\mzero}{\srzero}

\DeclareMathOperator{\oemptyset}{\varnothing}

\theoremstyle{definition}
\newtheorem{definition}{Definition}[section]

\newcommand{\CellWithForceBreak}[2][c]{
\begin{tabular}[#1]{@{}c@{}}#2\end{tabular}}

\newcommand\vldbdoi{XX.XX/XXX.XX}

\newcommand\vldbavailabilityurl{https://github.com/SparseLinearAlgebra/la-rpq}

\begin{document}
\title{Single-Source Regular Path Querying in Terms of Linear Algebra}

\author{Georgiy Belyanin}
\orcid{0009-0004-7527-4281}
\affiliation{%
  \institution{Saint-Petersburg State University}
  \city{Saint-Petersburg}
  \country{Russia}
}
\email{belyaningeorge@ya.ru}

\author{Rodion Suvorov}
\affiliation{%
  \institution{Saint-Petersburg State University}
  \city{Saint-Petersburg}
  \country{Russia}}
\email{suvorov.53245324@gmail.com}

\author{Semyon Grigorev}
\affiliation{%
  \institution{Saint-Petersburg State University}
  \city{Saint-Petersburg}
  \country{Russia}}
\orcid{0000-0002-7966-0698}
\email{s.v.grigoriev@mail.spbu.ru}

\begin{abstract}
    
\emph{Two-way regular path queries} (2-RPQs) allow one to use regular languages over edges and inverted edges in edge-labelled graph to constrain paths of interest. 2-RPQs are (partially) adopted in different real-world graph analysis systems and have become a part of the GQL ISO standard. However the performance of 2-RPQs on real-world graphs remains a bottleneck for wider adoption.
Utilisation of high-performance sparse linear algebra libraries for the algorithm implementation allows one to achieve significant speedup over competitors on real-world data and queries.

We propose a new breadth-first-search-based algorithm that leverages linear algebra for evaluating single-source regular path queries. We integrate it into the LAGraph graph processing algorithm infrastructure and provide in-depth performance comparison on the large real-world knowledge bases. Additionally, we present extensive analysis of its performance across different query types using synthetic data, comparing it with various databases and other linear algebra-based approaches.


\end{abstract}

%
%

\maketitle

\section{Introduction}

Language-constrained path querying~\cite{10.1137/S0097539798337716} is a way to search for paths in edge-labeled graphs where constraints are expressed in terms of a formal language.
The language restricts the set of valid paths: the sequence of labels along a path must form a sentence belonging to the language.
Regular languages are the most popular class of constraints used as navigational queries in graph databases.

Queries that employ regular languages to specify constraints are called \emph{regular path queries} or RPQs.
First introduced in 1989 by Alberto O. Mendelzon and Peter T. Wood~\cite{Mendelzon1989FindingRS}, RPQs have been intensively studied. 
Different extensions of RPQs are studying too: two-way RPQ or 2-RPQ that extends the alphabet of RPQs by the inverse of relationship symbols~\cite{calvanese2000query}, conjunctive RPQ or CRPQ that allows one to check several (parallel) paths~\cite{10.5555/3087011.3087032}.
Regular path constraints and their extensions have been (partially, in some cases) adopted in numerous graph analysis systems and query languages, including Cypher~\cite{10.1145/3183713.3190657} and PGQL~\cite{10.1145/2960414.2960421}. 
Moreover, RPQ is part of the ISO standard for the GQL graph query language~\cite{2024gql}, and the core of SPARQL 1.1 RDF query language~\cite{2013sparql}.

Despite their long history of theoretical and applied research, as well as real-world adoption, RPQs (and related extensions) remain in the focus of research.
One of the important directions is the implementation and optimization of RPQ evaluation algorithms~\cite{Bonifati2018} to achieve better performance in real world cases, and performance-targeted solutions are still actively developed~\cite{9835557, Arroyuelo2023, ma2024acceleratingregularpathqueries}.
Thus, designing new, efficient algorithms for RPQ evaluation remains an active challenge, as highlighted by Angela Bonifati et al in~\cite{Bonifati2018}.
Various approaches have been proposed to enhance RPQ evaluation performance, ranging from specialized indexing techniques~\cite{10.1007/978-3-030-85899-5_8, Arroyuelo2023} to parallel and distributed computing models~\cite{10.1145/2949689.2949711, 10.1145/3626562.3626833, Guo2021, Wang2019}.

Sparse linear algebra has emerged as a powerful paradigm for high-performance graph analysis, championed by the GraphBLAS community ~\cite{Kepner2016MathematicalFO}\footnote{GraphBLAS community web page: \url{https://graphblas.org/}}.
A vast number of graph analysis algorithms, such as PageRank or triangle centralities, can be expressed in terms of linear algebra\footnote{An almost complete list of graph analysis algorithms in terms of linear algebra: \url{https://github.com/GraphBLAS/GraphBLAS-Pointers}.} and the respective implementations demonstrate promising performance in real-world cases~\cite{9622789}.
Even more, it has been shown that sparse linear algebra enables a high-performance algorithm for more expressive query classes, such as Context-Free Path Queries (CFPQ)~\cite{Terekhov2021MultipleSourceCP}.
However, to our knowledge, there are only a few studies on linear-algebra-based RPQ algorithms.

On the one hand, the modern graph database FalkorDB\footnote{
Sources of FalkorDB on GitHub: \url{https://github.com/FalkorDB/falkordb}
} (formerly RedisGraph)~\cite{redisgraph} is based on SuiteSparse:GraphBLAS~\cite{10.1145/3577195}, reference implementation of the GraphBLAS API, and uses linear algebra for graph analysis.
While FalkorDB supports a subset of the Cypher query language, including some regular constraints, there is no detailed analysis of the respective algorithm.

At the same time, the linear-algebra-based RPQ evaluation algorithm was recently proposed by Diego Arroyuelo et al~\cite{10.1007/978-3-031-43980-3_4}. 
Despite using sparse matrices and parallel computations, this solution exhibits performance limitations, that are evident from the evaluation~\cite{10.1007/978-3-031-43980-3_4}. 

On the other hand, there are BFS-based strategies for the RPQ evaluation (e.g.~\cite{10.1007/978-3-642-31235-9_12})\footnote{For an overview of primary RPQ evaluation techniques, including relational algebra and automata-based approaches, and their optimization, we refer to the book ``Querying Graphs'' by Angela Bonifati et al~\cite{Bonifati2018}.}. 
Notably, BFS and its variants can be expressed using linear algebra~\cite{suitesparse}, suggesting an opportunity to develop a linear-algebra-based RPQ algorithm with BFS-like traversal at its core. We explore this direction in our work.

To summarize, in this work we make the following contributions.
\begin{enumerate}
    \item A novel BFS-based algorithm (LARPQ) for single-source (or symmetrically, single-destination) 2-RPQ is proposed. The algorithm is based on linear algebra: it is expressed in terms of operations over sparse boolean matrices. This fact allows one to utilize high-performance parallel libraries, such as SuiteSparse:GraphBLAS~\cite{10.1145/3577195}, for 2-RPQ evaluation. The correctness of the proposed algorithm is proven.
    \item Our implementation of the proposed algorithm, based on SuiteSparse:GraphBLAS, is evaluated and compared with other linear algebra-based solutions such as FalkorDB and the algorithm proposed by Diego Arroyuelo et al~\cite{10.1007/978-3-031-43980-3_4}, with graph databases Blazegraph and state-of-the-art MillenniumDB. Experimental results on real-world datasets (Wikidata with query logs from the MillenniumDB path query challenge~\cite{rpq-challenge} and Yago-2S) demonstrate that our solution achieves competitive performance. While occasionally slower on individual queries, our algorithm shows consistent average speedups: $6.8 \times$ for the algorithm of Diego Arroyuelo et al, $11.3 \times$ for MillenniumDB, $18.9 \times$ for FalkorDB, and $16.8 \times$ for Blazegraph. 
\end{enumerate}

\section{Preliminaries}

In this section, we provide the theoretical basics of graph theory and formal language theory required to define the RPQ problem and to describe our solution and the algorithm proposed by Diego Arroyuelo et al.

First we define the edge-labeled graph that we use as a data model.
\begin{definition}[Edge-labelled graph]
    A quadruple \\ $\mathcal{G} = \langle V, E, L, \lambda_\mathcal{G} \rangle$ is called an {\em edge-labelled graph} (or a {\em graph}) if:
    \begin{itemize}
        \item $V$ is a finite set of vertices;
        \item $E \subseteq V \times V$ is a finite set of edges;
        \item $L$ is a finite set of labels;
        \item $\lambda_{\mathcal{G}}: E \mapsto 2^L$ represents edge labels.
    \end{itemize}
\end{definition}

Any finite set can be enumerated by natural numbers from $1$ to $n$. For the rest of the paper, we will assume that the vertices of the graph $\mathcal{G} = \langle V, E, L, \lambda_\mathcal{G} \rangle$ are enumerated, and without loss of generality $V = \{ 1, 2, ..., |V| \}$.

Let $\mathcal{G} = \langle V, E, L, \lambda_{\mathcal{G}} \rangle$ be an edge-labelled graph. Introduce symbols and sets:
\begin{itemize}
    \item $a^- \notin L$ for $a \in L$, $(a^-)^- = a$ and $a = b \Leftrightarrow a^- = b^-$
    \item $e^- = (v, u)$ where $e = (u, v)$.
    \item $E^- = \{e^- \mid e \in E\}$.
    \item $L^- = \{a^- \mid a \in L\}$.
    \item $\lambda_{\mathcal{G}}^-: E^- \mapsto 2^{L^-}, \; \lambda_{\mathcal{G}}^-(e^-) = \{ a^- \mid a \in \lambda_\mathcal{G}(e) \}$.
\end{itemize}

We also need these sets to generalize the definition of the directed graph to be able to traverse it in both directions:

\begin{itemize}
    \item $E^\leftrightarrow = E \cup E^-$.
    \item $L^\leftrightarrow = L \cup L^-$.
    \item $\lambda_{\mathcal{G}}^\leftrightarrow: E^\leftrightarrow \mapsto 2^{L^\leftrightarrow},\;\; \lambda_{\mathcal{G}}^\leftrightarrow(e) = \lambda_\mathcal{G}(e) \cup \lambda^-_\mathcal{G}(e) \;$.
    \item $\mathcal{G}^\leftrightarrow = \langle V, E^\leftrightarrow, L^\leftrightarrow, \lambda^\leftrightarrow \rangle$.
\end{itemize}

\begin{definition}[path]
A {\em path} $\pi = (e_1, e_2, ..., e_n)$ of length $n$ in the graph $\mathcal{G} = \langle V, E, L, \lambda_{\mathcal{G}} \rangle$ from $u_1$ to $v_n$ is a finite sequence of edges $e_i = (u_i, v_i) \in E$ s.t. $\forall 1 \le i \le n - 1 \; v_i = u_{i+1}$.

We say there are {\em zero-length paths} represented by an empty sequence from $v$ to $v$ for all $v \in V$.
\end{definition}

\begin{definition}[2-way path]
A {\em 2-way path} (2-path) $\pi = (e_1, e_2, ..., e_n)$ in the graph $\mathcal{G} = \langle V, E, L, \lambda_{\mathcal{G}} \rangle$ is a path in $\mathcal{G}^\leftrightarrow = \langle V, E^\leftrightarrow, L^\leftrightarrow, \lambda_{\mathcal{G}}^\leftrightarrow \rangle$.
\end{definition}

The $\omega$ maps paths to words as defined below.
\begin{itemize}
    \item $\omega_\mathcal{G}(\pi) = \{ a_1 \cdot a_2 \cdot ... \cdot a_n \mid a_i \in \lambda_\mathcal{G}(e_i) \}$ for path $\pi = (e_1, e_2, ..., e_n)$ in $\mathcal{G}$.
    \item $\omega^\leftrightarrow_\mathcal{G}(\pi) = \{ a_1 \cdot a_2 \cdot ... \cdot a_n \mid a_i \in \lambda^\leftrightarrow_\mathcal{G}(e_i) \}$ for 2-path $\pi = (e_1, e_2, ..., e_n)$ in $\mathcal{G}$.
\end{itemize}
 where $\cdot$ denotes concatenation.


%
%
%

Regular languages (RLs) represent the set of all languages that are accepted by finite automata. However, we are going to look for 2-way paths for which we need an NFA modification to be introduced.

\begin{definition}[2-way non-deterministic finite automaton]
    A {\em2-way non-deterministic finite automaton} (2-NFA) is a tuple $\mathcal{N} = \langle Q, \Sigma, \Delta, \lambda_{\mathcal{N}}, Q_S, Q_F\rangle$, where:
    \begin{itemize}
        \item $Q$ is a finite set of states;
        \item $\Sigma$ is a finite alphabet;
        \item $\Delta \subseteq Q \times Q$ is transition relation;
        \item $\lambda_{\mathcal{N}}: \Delta \mapsto 2^{\Sigma^{\leftrightarrow}}$ assigns a set of labels (including inverses) to each transition;
        \item $Q_S \subseteq Q$ is a set of starting states;
        \item $Q_F \subseteq Q$ is a set of final states.
    \end{itemize}
\end{definition}


The set of languages accepted by 2-NFAs coincides with the set of all regular languages over $\Sigma^{\leftrightarrow}$. Let $[\![ \mathcal{N} ]\!] = \mathcal{R}$ where $\mathcal{R}$ is the RL accepted by the 2-NFA $\mathcal{N}$.

Notice that 2-NFA can be seen as a graph $\mathcal{G}_\mathcal{N} = \langle Q, \Delta, \Sigma^{\leftrightarrow}, \lambda_\mathcal{N} \rangle$ equipped with the set of starting states $Q_S \subseteq Q$ and the set of final states $Q_F \subseteq Q$. Analogously introduce sets $\Delta^-$, $\lambda_\mathcal{N}^-$ and the map $\omega_{\mathcal{N}}(\pi) = \{ a_1 \cdot a_2 \cdot ... \cdot a_n \mid a_i \in \lambda_{\mathcal{N}}(\delta_i) \}$ for the paths $\pi = (\delta_1, \delta_2, ..., \delta_n)$ in the graph $\mathcal{G}_\mathcal{N}$.

Conversely, a graph $\mathcal{G} = \langle V, E, L, \lambda_\mathcal{G} \rangle$ equipped with some $V_S \subseteq V$ can be seen as 2-NFA $\mathcal{N}_\mathcal{G} = \langle V, L, E, \lambda_\mathcal{G}^\leftrightarrow, V_S, V \rangle$. Such 2-NFA accepts the language:
\[
[\![ \mathcal{N}_G ]\!] = \bigcup_{\substack{\text{2-path } \pi_\mathcal{G} \text{ in } \mathcal{G} \\ \text{ from } v_S \in V_S}} \omega_\mathcal{G}^\leftrightarrow(\pi_\mathcal{G})
\]
Thus, we can treat the evaluation of 2-RPQs with a fixed set of starting vertices as an intersection of 2-NFAs. 
\begin{definition}[2-NFA intersection]
    For two arbitrary 2-NFAs $\mathcal{N}_1 = \langle Q_1, \Sigma_1, \Delta_1, \lambda_{\mathcal{N}_1}, Q_{S1}, Q_{F1}\rangle$ and $\mathcal{N}_2 = \langle Q_2, \Sigma_2, \Delta_2, \lambda_{\mathcal{N}_2}, Q_{S2}, Q_{F2}\rangle$ introduce a new automaton $\mathcal{N} = \mathcal{N}_1 \times \mathcal{N}_2$ called the {\em intersection of 2-NFAs $\mathcal{N}_1$ and $\mathcal{N}_2$} where $\mathcal{N} = \langle Q, \Sigma, \Delta, \lambda_{\mathcal{N}}, Q_S, Q_F\rangle$ s.t.:
    \begin{itemize}
        \item $Q = Q_1 \times Q_2$;
        \item $\Sigma = \Sigma_1 \cap \Sigma_2$;
        \item $\Delta = \{ ((q_1, q_2), (q_1', q_2')) \mid (q_1, q_1') \in \Delta_1$, $(q_2, q_2') \in \Delta_2 \}$;
        \item $\lambda_{\mathcal{N}}(((q_1, q_2), (q_1', q_2'))) = \lambda_{\mathcal{N}_1}((q_1, q_1')) \cap \lambda_{\mathcal{N}_2}((q_2, q_2'))$;
        \item $Q_S = Q_{S1} \times Q_{S2}$;
        \item $Q_F = Q_{F1} \times Q_{F2}$.
    \end{itemize}
\end{definition}

The resulting 2-NFA accepts the intersection of the languages defined by the initial automata $\mathcal{N}_1$ and $\mathcal{N}_2$: 
\[
[\![ \mathcal{N}_1 \times \mathcal{N}_2 ]\!] = [\![ \mathcal{N}_1 ]\!] \cap [\![ \mathcal{N}_2 ]\!].
\] 


We have all the necessary preliminaries to formally state the problem solved by the 2-RPQ algorithm.
\begin{definition}[Two-way regular path query]
    Recall {\em two-way regular path query} (2-RPQ) a 4-tuple $\langle \mathcal{G}, \mathcal{R}, V_s, V_f\rangle$ where $\mathcal{G} = \langle V, E, L, \lambda_{\mathcal{G}} \rangle$ is a graph, $\mathcal{R}$ is a regular language over the alphabet $L^\leftrightarrow$, $V_s \subseteq V$ is a set of starting nodes, and  $V_f \subseteq V$ is a set of final nodes.
\end{definition}
Frequent practical cases are queries with fixed starting or final vertex. So, for 2-RPQ $\mathcal{Q} = \langle \mathcal{G}, \mathcal{R}, V_{s}, V_f\rangle$ we are interested in an efficient way of evaluating these maps:
\begin{itemize}
    \item {\bf Single-source reachability} ($V_s = \{ v_s \}, V_f = V$ ):
    \[
        [\![ \mathcal{Q} ]\!]_{SSR} = 
        \left\{\;
            u \in V \;\middle|\;
            \begin{gathered}
                \exists \text{ 2-path } \pi \text{ in } \mathcal{G} \text{ from } v_{s} \text{ to } u \\ \text{and }
                \omega^\leftrightarrow_{\mathcal{G}}(\pi) \cap \mathcal{R} \ne \oemptyset\\
            \end{gathered}
        \;\right\}
    \]
    \item {\bf Single-destination reachability} ($V_s = V , V_f = \{ v_f \}$):
    \[
        [\![ \mathcal{Q} ]\!]_{SDR} = 
        \left\{\;
            u \in V \;\middle|\;
            \begin{gathered}
                \exists \text{ 2-path } \pi \text{ in } \mathcal{G} \text{ from } u \text{ to } v_{f} \\ \text{and }
                \omega^\leftrightarrow_{\mathcal{G}}(\pi) \cap \mathcal{R} \ne \oemptyset\\
            \end{gathered}
        \;\right\}
    \]
\end{itemize}

For such partial cases we introduce the following short versions: $Q_{SSR} = \langle \mathcal{G}, \mathcal{R}, v_s\rangle$ and $Q_{SDR} = \langle \mathcal{G}, \mathcal{R}, v_f\rangle$ respectively.

\section{Linear Algebra, Graphs and Relations}


In order to operate with graphs in terms of linear algebra we need to see how algebraic objects are connected with set relations and how the relations are connected to graph theory.

For two enumerated finite sets $A = \{ 1, 2, ... n \}$, $B = \{1, 2, ..., m\}$ for some $n, m \in \mathbf{N}$ a binary matrix $T$ of size $|A| \times |B|$ can be used to represent a relation $\mathcal{T} \subseteq A \times B$: $T_{ij} = 1 \Leftrightarrow (i, j) \in \mathcal{T}$. Recall this matrix $T$ {\em a matrix representing the relation } $\mathcal{T}$.

We need to define the following linear algebra operations over the Boolean matrices. Let $A$, $B$, $C$ be the matrices over Boolean algebra $\langle \mathcal{B} = \{0, 1\}, \vee, \wedge, \neg, 0, 1 \rangle$. Introduce the following operations and constants.


\begin{definition}
    \emph{Zero matrix} is a rectangular matrix $\mzero_{n \times k}$, such that $\mzero_{ij} = 0$ for all valid $i$ and $j$.
\end{definition}

\begin{definition}
    For the Boolean matrix $A_{n \times m}$, the matrix $B_{n \times m}$, such that $B_{ij} = \neg A_{ij}$ is a \emph{complement matrix} for the matrix $A$. We denote complementation of $A$ as $\neg A$.
\end{definition}
    
\begin{definition}
    For the Boolean matrix $A_{n \times m}$, the matrix $B_{m \times n}$, such that $B_{ij} = A_{ji}$ is a \emph{transposed matrix} for the matrix $A$. We denote transposition of $A$ as $A^T$.
\end{definition}

\begin{definition}
    For the given Boolean matrices $A_{n \times m}$ and $B_{n \times m}$, the \emph{sum} $A \oplus B$ is a matrix $C_{n \times m}$ such that matrix $C_{ij} = A_{ij} \vee B_{ij}$.
\end{definition}

\begin{definition}
    For the given Boolean matrices $A_{n \times m}$ and $B_{m \times k}$, the \emph{product} $A \otimes B$ is a matrix $C_{n \times k}$ such that matrix $$C_{ij} = \bigvee_{l=1}^m A_{il} \wedge B_{lj}.$$
\end{definition}

\begin{definition}
    For the given Boolean matrices $A_{n \times m}$ and $B_{n \times m}$, the \emph{masking} $A \langle B \rangle$ is a matrix $C_{n \times m}$ such that matrix $C_{ij} = A_{ij} \wedge B_{ij}.$
\end{definition}



Let $\mathcal{A}$, $\mathcal{B}$ and $\mathcal{C}$ be binary relations over sets $S_1$, $S_2$, $S_3$. Let $A$, $B$ and $C$ be binary matrices representing them. Then the following correspondences hold.
\begin{itemize}
    \item 
        \begin{tabular}[t]{p{1.3cm}l}
            $C = A \mplus B$ & $\leftrightarrow \mathcal{C}=\mathcal{A} \cup \mathcal{B}$\\
            &\;\;\;\;\;where $\mathcal{A}, \mathcal{B}, \mathcal{C} \subseteq S_1 \times S_2$.
        \end{tabular}
    \item
        \begin{tabular}[t]{p{1.3cm}l}
            $C = A \mtimes B$ & $\leftrightarrow \mathcal{C}=\{ (i, k) \mid (i, j) \in \mathcal{A}$, $(j, k) \in \mathcal{B} \}$ \\
            &\;\;\;\;\;if $\mathcal{A} \subseteq S_1 \times S_2$, $\mathcal{B} \subseteq S_2 \times S_3$ and \\
            &\;\;\;\;\;$\mathcal{C} \subseteq S_1 \times S_3$.
        \end{tabular}
    \item
        \begin{tabular}[t]{p{1.3cm}l}
            $C = A^T$ & $\leftrightarrow \mathcal{C} = \{ (i, j) \mid (j, i) \in \mathcal{A} \} = \mathcal{A}^-$ \\
            &\;\;\;\;\;where $\mathcal{A} \subseteq S_1 \times S_2$, $C \subseteq S_2 \times S_1$.
        \end{tabular}
    \item 
        \begin{tabular}[t]{p{1.3cm}l}
        $C = A \langle B \rangle$ & $\leftrightarrow \mathcal{C} = \mathcal{A} \cap \mathcal{B}$, \\
        &\;\;\;\;\;where $\mathcal{A}, \mathcal{B}, \mathcal{C} \subseteq S_1 \times S_2 $.
        \end{tabular}
    \item
        \begin{tabular}[t]{p{1.3cm}l}
            $C = \neg A$ & $\leftrightarrow \mathcal{C} = \{ (i, j) \mid (i, j) \in S_1 \times S_2 \setminus \mathcal{A} \}$ \\
            &\;\;\;\;\;where $\mathcal{A}, \mathcal{C} \subseteq S_1 \times S_2$.
        \end{tabular}
\end{itemize}


For a given arbitrary graph $\mathcal{G} = \langle V, E, L, \lambda_\mathcal{G} \rangle$ and 2-NFA $\mathcal{N} = \langle Q, \Sigma, \Delta, \lambda_{\mathcal{N}}, Q_S, Q_F \rangle$ introduce the sets:
\begin{itemize}
    \item $E^a = \{ e \mid e \in E, a \in \lambda^\leftrightarrow_{\mathcal{G}}(e) \}$ for all $a \in L^\leftrightarrow$.
    \item $\Delta^a = \{ \delta \mid \delta \in \Delta, a \in \lambda_{\mathcal{N}}(\delta) \}$ for all $a \in \Sigma^\leftrightarrow$.
\end{itemize}

These sets consist of the edges of the graph $\mathcal{G}$ and the transitions of 2-NFA $\mathcal{N}$ marked with a label $a$. They can be seen as binary relations over the sets $V$ and $Q$ correspondingly.

\begin{definition}[Adjacency matrix of the label]
    Let $G^a$ be the matrix representing the binary relation $E^a$. $G^a$ is called {\em an adjacency matrix of the label $a$}.
\end{definition}

\begin{definition}[Boolean decomposition of the adjacency matrix]
    A {\em Boolean decomposition of the adjacency matrix } $G$ is a set of Boolean matrices $\{ G^{a} \mid a \in L\}$. 
\end{definition}

Boolean decomposition of the adjacency matrices of some real-world data represented by graphs is a set of \textbf{sparse} matrices. This fact strictly leads to the idea of exploiting it for an efficient representation and using sparse matrix operation algorithm implementations.

Also note that for the given graph $\mathcal{G} = \langle V, E, L, \lambda_\mathcal{G} \rangle$, $E^{a^-} = (E^a)^T$ and $G^{a^-} = (G^a)^T$.

\section{BFS-Based Single-Source RPQ in Terms of Linear Algebra}

\newcommand\mycommfont[1]{\footnotesize\ttfamily\textcolor{blue}{#1}}
\SetCommentSty{mycommfont}

In this section we describe a single-source linear-algebra based 2-RPQ algorithm and prove its correctness. 

\begin{algorithm}
    \caption{Single Source 2-RPQ using linear algebra}
    \label{ssr-algorithm}
    \SetKwInOut{Input}{input}
    \SetKwInOut{Output}{output}
    \Input{$\mathcal{N} = \langle Q, \Sigma, \Delta, \lambda_\mathcal{N} Q_S, Q_F \rangle$, $\mathcal{G}=\langle V, E, L, \lambda_\mathcal{G} \rangle$, $v_s \in V$}
    \Output{Vector $P^F$ of size $1 \times |V|$}
    
    $\{ N^a \}\gets$ Boolean decomposition of the $\mathcal{N}$ adjacency matrix\;
    $\{ G^a \}\gets$ Boolean decomposition of the $\mathcal{G}$ adjacency matrix\;
    $P_{|Q| \times |V|} \gets \mzero_{|Q| \times |V|}$ \;
    $M_{|Q| \times |V|} \gets M $ s.t. $ M_{qv} = 1$ if $q \in Q_S$, $v = v_s$, otherwise $0$\;
    $F_{1 \times |Q|} \gets F $ s.t. $F_1q = 1$ if $q \in Q_F$, otherwise $0$\;
    \While{$M \ne \mzero$}{
        $\displaystyle M \gets \bigoplus_{a \in \Sigma^\leftrightarrow \cap L^\leftrightarrow}((N^a)^T \mtimes M \mtimes G^a) \langle \neg P \rangle$\tcp*{Update $\mathcal{M}$}
        $P\gets P \mplus M$
    }
    
    \Return{$P^F = F \mtimes P$}
\end{algorithm}

Let $\mathcal{N} = \langle Q, \Sigma, \Delta, \lambda_{\mathcal{N}}, Q_S, Q_F \rangle$ be the input 2-NFA that represents the query and specifies the RL $\mathcal{R} = [\![ \mathcal{N} ]\!]$. Let $\mathcal{G}=\langle V, E, L, \lambda_{\mathcal{G}} \rangle$ be the input graph and $v_{s} \in V$ is a starting node. Then the algorithm \ref{ssr-algorithm} 
builds the following automata intersection:

\[
    [\![ \mathcal{N} \times \mathcal{N}_G ]\!] = [\![ \mathcal{N} ]\!] \cap [\![ \mathcal{N}_\mathcal{G} ]\!] = 
    \mathcal{R} \cap \left(\bigcup_{\substack{\text{2-path } \pi_\mathcal{G} \text{ in } \mathcal{G} \\ \text{ from } v_S \in V_S}} \omega_\mathcal{G}^\leftrightarrow(\pi_\mathcal{G})\right).
\]

Result of the algorithm is a vector $P^F_{1 \times |V|}$ for which $v \in [\![ \mathcal{Q} ]\!]_{SSR} \Leftrightarrow P^F_{1v} = 1$ for some 2-RPQ $\mathcal{Q} = \langle \mathcal{G}, \mathcal{R}, v_s \rangle$.

The core of the algorithm is line 8 that performs one step of traversing two automata simultaneously and builds matrix $P_{|Q| \times |V|}$ such that: 

\[
\begin{cases}
P_{q_Fv} = 1 \\
q_F \in Q_F
\end{cases}
\Leftrightarrow 
\begin{cases}
    \exists \pi_\mathcal{G} \text{ in } \mathcal{G} \text{ from } v_s \text { to } v \\
    \omega^\leftrightarrow_\mathcal{G}(\pi_\mathcal{G}) \cap \mathcal{R} \neq \oemptyset.
\end{cases}
\]

\begin{figure}[h]
    \includegraphics[width=0.33\textwidth]{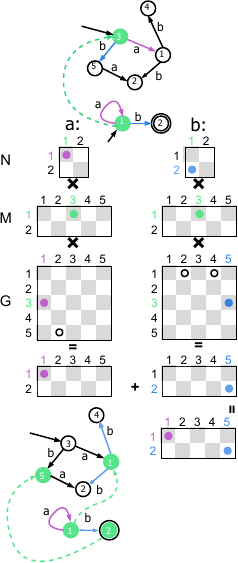}
    \caption{The algorithm step for graph and automaton for regular expression $a^*b$}
    \label{algorithm-step}
\end{figure}

An example of such a step is presented in figure~\ref{algorithm-step}.
For graph and automaton with labels $\{a,b\}$, we visualize adjacency matrices for both of these symbols, the matrix of relation $M$, and the process of new $M$ computation.
The visualization uses solid-colored edges to represent transitions traversed simultaneously at the step, while dashed edges show the reachability relation $M$ --- connecting vertices that are simultaneously reachable.
Initially, $M$ contains just one edge linking the automaton's initial state to the starting vertex in the graph.
After one iteration of the main cycle, $M$ contains two edges.
One of them connects the final state of the automaton with vertex 5 of the graph.
This indicates that vertex 5 is reachable from the starting vertex 3 by the path that forms a word acceptable by the automaton.
%

%

The core idea of the algorithm can be summarized in a theorem that is proved using straightforward induction by the length of the paths. Details are provided in appendix~\ref{appendix:proof-of-correctness}. 

\begin{theorem}[LA 2-RPQ algorithm correctness] 
The proposed algorithm, represented in~\ref{ssr-algorithm}, computes the matrix $P$ such that the respective relation $\mathcal{P} \subseteq Q \times V$ has the following property.
\[
(q, v) \in \mathcal{P} \Leftrightarrow 
\begin{cases}
    \exists \text{ 2-path } \pi_G \text{ in } \mathcal{ G } \text{ from } v_s \text{ to } v \\
    \exists \text{ path } \pi_N \text{ in } \mathcal{N} \text{ from some } q_s \in Q_S \text{ to } q \\
    \omega^\leftrightarrow_\mathcal{G}(\pi_\mathcal{G}) \cap \omega_\mathcal{N}(\pi_\mathcal{N}) \ne \oemptyset.
\end{cases}
\]
\end{theorem}

In particular, from the theorem we can immediately conclude that for the given automaton $\mathcal{N}$, graph $\mathcal{G}$ and starting vertex $v_s$, $P_{q_Fv} = 1$ for $q_F \in Q_F$ iff exist two paths $\pi_\mathcal{G}$ to $v$ and $\pi_\mathcal{N}$ to $q_F$ such that $\omega^{\leftrightarrow}_{\mathcal{G}}(\pi_\mathcal{G}) \cap \omega_\mathcal{N}(\pi_\mathcal{N}) \neq \oemptyset$. 
As far as $q_F$ is a final state of the NFA, $\omega^{\leftrightarrow}_{\mathcal{G}}(\pi_\mathcal{G}) \subseteq [\![ \mathcal{N} ]\!]$. Thus, $v$ reachable from $v_s$ by path satisfies the given regular constraint.

The proposed single-source 2-RPQ algorithm can be used to solve the single-destination problem. Let $\mathcal{N} = \langle Q, \Sigma, \Delta, \lambda_{\mathcal{N}}, Q_S, Q_F \rangle$ be an NFA specifies the constraint in the single destination query. Denote: 
\begin{itemize}
    \item $\mathcal{R}^- = \{ w^- = a_n^-a_{n-1}^-...a_2^-a_1^- \mid w = a_1a_2...a_n) \in \mathcal{R} \}$
    \item $\mathcal{N}^- = \langle Q, \Sigma, \Delta^-, \lambda_{\mathcal{N}}^-, Q_F, Q_S \rangle$.
    \item $\mathcal{Q}^- = \langle \mathcal{G}, \mathcal{R}^-, v_{s} \rangle$.
\end{itemize} 


The automaton $\mathcal{N}^-$ allows one to traverse paths in the opposite direction. It holds $\mathcal{R}^- = [\![ \mathcal{N}^- ]\!]$. Evaluating single-source reachability 2-RPQ on the reversed 2-RPQ $\mathcal{Q}^-$ gives using the 2-NFA $\mathcal{N}^-$:
\[
[\![\mathcal{Q}]\!]_{SDR}=[\![\mathcal{Q}^-]\!]_{SSR}.
\]
Boolean matrix decomposition of $\mathcal{N}^-$ can be easily obtained by transposing the matrices in Boolean matrix decomposition of $\mathcal{N}$.


For performance reasons, it is necessary to take into account the fact that sparse matrix multiplication works faster when matrices have fewer nonzero entries. The associativity of the Boolean matrix multiplication can be used to perform less calculations. This leads to the idea of two possible ways to calculate the product on line 8 in the algorithm \ref{ssr-algorithm}: 
$((N^a)^T \mtimes M) \mtimes G^a = (N^a)^T \mtimes (M \mtimes G^a).
$

In general, the first option is preferable due to the fact that 2-NFA for the query can be converted into the corresponding minimal deterministic finite automaton having no more than $|Q|$ non-zero cells in each $\mathcal{N}^a$ adjacency matrix due to the automaton determinism. This strictly leads to a less expensive computation of the product with the large $|V| \times |V|$ matrix.

When it comes to real-world implementation, it is important to note that it is also possible to implement BFS-based 2-RPQs without using a traversal matrix $M$ as described in algorithm \ref{ssr-algorithm-no-mask}. The complete description of such algorithm is available in appendix \label{appendix:no-mask-algorithm}. For some sparse linear algebra libraries, it can be more efficient to deal with fewer distinct matrices than to have fewer nonzero entries, especially when there are very few of them. Such properties are common for executing simple regular path queries.

\[
(q, v) \in \mathcal{P}_n \Leftrightarrow
\begin{cases}
    \exists \text{ 2-path } \pi_\mathcal{G} \text{ of length } \leq n \text{ in } \mathcal{ G } \text{ from } v_s \text{ to } v \\
    \exists \text{ path } \pi_\mathcal{N} \text{ of length } \leq n \text{ in } \mathcal{N} \text{ from } q_s \in Q_S \text{ to } q \\
    \omega^\leftrightarrow_\mathcal{G}(\pi_\mathcal{G}) \cap \omega_\mathcal{N}(\pi_\mathcal{N}) \ne \oemptyset.
\end{cases}
\]

\begin{algorithm}
    \caption{Single Source 2-RPQ using linear algebra without an extra traversal matrix}
    \label{ssr-algorithm-no-mask}
    \SetKwInOut{Input}{input}
    \SetKwInOut{Output}{output}
    \Input{$\mathcal{N} = \langle Q, \Sigma, \Delta, \lambda_\mathcal{N} Q_S, Q_F \rangle$, $\mathcal{G}=\langle V, E, L, \lambda_\mathcal{G} \rangle$, $v_s \in V$}
    \Output{Vector $P^F$ of size $1 \times |V|$}
    
    $\{ N^a \}\gets$ Boolean decomposition of the $\mathcal{N}$ adjacency matrix\;
    $\{ G^a \}\gets$ Boolean decomposition of the $\mathcal{G}$ adjacency matrix\;
    $P_{|Q| \times |V|} \gets P $ s.t. $ P_{qv} = 1$ if $q \in Q_S$, $v = v_s$, otherwise $0$\;
    $F_{1 \times |Q|} \gets F $ s.t. $F_1q = 1$ if $q \in Q_F$, otherwise $0$\;

    \While{$P$ changes}{
        $\displaystyle P \gets P \oplus \bigoplus_{a \in \Sigma^\leftrightarrow \cap L^\leftrightarrow}((N^a)^T \mtimes P \mtimes G^a)$\tcp*{Update $\mathcal{P}$}
    }
    
    \Return{$P^F = F \mtimes P$}
\end{algorithm}

Although it may seem that this algorithm should perform some extra calculations because $P$ contains more entries than $M$ on each step within algorithm \ref{ssr-algorithm}, this way of implementing the proposed algorithm can be handy. Some sparse linear algebra libraries perform better when handling fewer distinct matrices, even if that increases the total number of non-zero entries. Such properties are common for executing simple regular path queries.

\subsection{Comparison With Diego Arroyuelo Algorithm}

Diego Arroyuelo et al. in~\cite{10.1007/978-3-031-43980-3_4} propose another linear algebra-based 2-RPQ evaluation algorithm called RPQ-matrix. It directly translates 2-RE to Boolean matrix operations rather than evaluating step-by-step traversal over the graph and the automaton with capability to employ single-source and single-destination 2-RPQs. The algorithm consists of the following steps.

\begin{enumerate}
    \item Build an abstract syntax tree (evaluation plan) representing the desired regular expression in which leaves represent labels and other nodes represent operations such as concatenation, Kleene star, and conjunction.
    \item Match nodes with matrices: each leaf label is matched with an adjacency matrix representing it, and every inner node is matched with a matrix that can be computed based on the child matrices depending on the operation the node represents.
    \item Optimize the evaluation plan by using the provided source or destination vertex.
    \item Compute the matrix representing the root element with operations reordering applied to optimize computations.
\end{enumerate}

We conduct experiments on several graph database management systems and compare them to RPQ-matrix and to the proposed algorithm in order to investigate the efficiency of different linear algebra-based approaches.




\section{Evaluation}

The proposed single-source 2-RPQ algorithm is implemented using the SuiteSparse:GraphBLAS library \cite{10.1145/3322125} within the LAGraph \cite{lagraph} infrastructure\footnote{Regular Path Query algorithm in LAGraph repository: \url{https://github.com/GraphBLAS/LAGraph/blob/stable/experimental/algorithm/LAGraph_RegularPathQuery.c}}.
Both sparse matrices and their transpositions are loaded in memory due to the fact that we need both representations to efficiently evaluate 2-RPQ and traverse the graph in both directions.

Whereas we are focusing on an efficient way of evaluating RPQs in memory, there are not many candidates to compare. Popular database management systems primarily focus on general availability and ensure availability of the concurrent access at the same time the suggested algorithm implementation is suited only for solving the reachability problem. We have selected the following systems as the most effective and the most related to our use case.

{\bf RPQ-matrix}~\cite{10.1007/978-3-031-43980-3_4} is another implementation of the linear algebra-based RPQ evaluation algorithm. The original work introduces a few variations of the adjacency matrix representation: $k^2$-trees offering less memory consumption and CSR/CSC formats providing better performance. We have chosen the last one to compare since we are aiming to compare the performance.

{\bf RPQ-matrix (GrB)}\footnote{SuiteSparse:GraphBLAS-based implementation of RPQ-matrix: \url{https://github.com/suvorovrain/rpq-matrix/tree/gbmod}} is an adapted version of Diego Arroyuelo et al. RPQ-matrix algorithm where matrix representations and operations are substituted with SuiteSparse:GraphBLAS equivalents in order to analyse performance impact of basic primitives implementation.

{\bf MillenniumDB}~\cite{millenniumdb} is a graph-oriented database management system with RDF-model and SPARQL support. It supports a synthetic way to carry out calculations without using disk storage by caching query data. MillenniumDB demonstrates state-of-the-art performance on evaluating regular path queries.

{\bf FalkorDB} (previously RedisGraph~\cite{redisgraph}) is an in-memory property graph database that also employs  SuiteSparse:GraphBLAS for query evaluation. However, it uses OpenCypher modification and supports only a subset of regular path queries (e.g., it is impossible to evaluate repeated-path queries in the form of $(a\;b)^*$). To deal with this, we have carried out the measurements of cypher-compatible and non-cypher-compatible queries separately.

{\bf Blazegraph}~\cite{millenniumdb} is another graph-oriented database management system using RDF data model and SPARQL query language. It is used by the Wikidata project and is based on more classical B-trees.

All experiments are conducted on a work station with Ryzen 9 7900X 4.7 GHz 12-core, 128 Gb of DDR5 RAM and running Ubuntu 22.04.

\subsection{Implementation Details}

As mentioned above, it is important for better performance to take into account that there are two different possible implementations of the BFS-based 2-RPQ algorithm: one involving less different matrices and one with less dense matrices. For SuiteSparse:GraphBLAS, the first approach turns out to be faster if the number of entries in matrices is small, and the latter is better for denser matrices. 

The most straightforward way to combine these two approaches to achieve better performance is to switch from the first approach to the second one during the traversal if the resulting matrix starts having some constant amount of non-zero entries. This constant can be empirically determined for the graph\footnote{For the studied datasets the most suitable value is $100$.}. This allows the BFS-based algorithm to provide strong performance on simple queries with very few answers and on analytical queries involving a lot of resulting vertices at the same time.

\subsection{Dataset Description}

To compare performance of different approaches we start from evaluating benchmarks on large real-world datasets. For algorithm evaluation, we choose the Wikidata dataset from the snapshot provided in terms of MillenniumDB path query challenge \cite{rpq-challenge}. The resource contains both the graph and the set of 660 different 2-RPQs in SPARQL format taken from the Wikidata query log. The second dataset is Yago-2S evaluated with 7 different complex queries taken from~\cite{yago-queries}.

We also want to compare different linear algebra-based approaches and determine the best of them for different query kinds. Real-world datasets are not suitable for it due to complex graph topology. Instead, we employ the synthetic RPQBench dataset generator~\cite{rpqbench} for controlled performance evaluation across query categories. The structure of the graph ensures reproducible and predictable results when similar queries are executed. The original generator produces arbitrary sized RDF datasets and offers $10$ different query kinds without starting or final nodes specified in SPARQL format. We generate a graph and supply these query kinds with randomly generated source and destination vertices.

For systems that do not support the RDF format, the datasets have been converted to an edge-labeled graph. Original SPARQL path queries have been deprefixed, converted to the corresponding minimized DFAs. Queries have been converted to Cypher queries of the form {\tt MATCH ... COUNT (DISTINCT ...)} when it is possible. Queries without starting or final vertex, broken queries involving missing entities have been removed. Final dataset statistics can be summarised as follows.

\noindent\textbf{Wikidata}
\begin{itemize}
    \item $610$ million edges, $91$ million vertices, $1400$ distinct labels.
    \item $578$ queries filtered out from $660$ from the query dump.
\end{itemize}

\noindent\textbf{Yago-2S}
\begin{itemize}
    \item $610$ million edges, $91$ million vertices, $1400$ distinct labels.
    \item $7$ complex queries taken from \cite{yago-queries}.
\end{itemize}

\noindent\textbf{RPQBench}
\begin{itemize}
    \item Synthetic dataset, $610$ million edges, $91$ million vertices, $1400$ distinct labels.
    \item $20$ different query kinds supplied $1000$ randomly generated source/destination vertices.
    \item Trivial queries are filtered out (e.g. $a^*$ has $\geq 1$ answer).
\end{itemize}

\subsection{Evaluation Scenario}

\begin{figure}
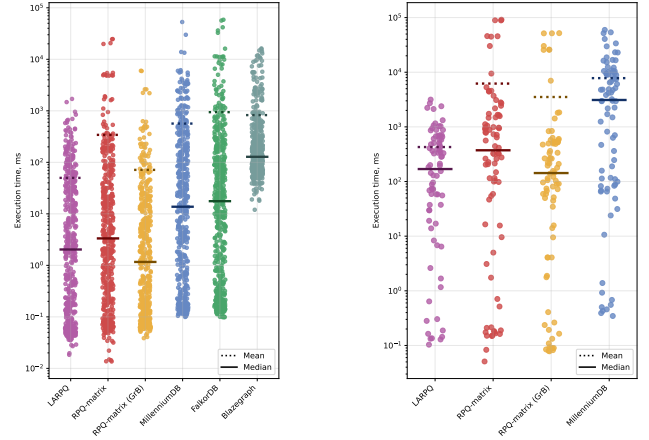

    \centering
    \begin{subfigure}[b]{0.21\textwidth}
        \centering
        \includesvg[width=\textwidth]{plots/wikidata-all.svg}
        \caption{Queries which are supported and succeeded on all of the competitors}
        \label{all-results}
    \end{subfigure}
    \hfill
    \begin{subfigure}[b]{0.21\textwidth}
        \centering
        \includesvg[width=\textwidth]{plots/wikidata-complex.svg}
        \caption{Queries which are not supported in Cypher or timed out on FalkorDB or Blazegraph}
        \label{complex-results}
    \end{subfigure}
    \caption{Wikidata dataset per-query evaluation time}
\end{figure}

\begin{table*}[t]
\centering
\caption{Wikidata query execution results. The results are supplied for simple and for complex (C) queries separately. For in-memory algorithms the memory consumption for whole dataset and bytes-per-triple values (BPT) are presented. Mean and median speedup relatively the proposed solution is a ratio of mean and median over query set for respective systems}
    \label{all-results-table}
\begin{tabular}{|l|c|c|c|c|c|c|}
\hline
 & LARPQ & RPQ-matrix & RPQ-matrix (GrB) & MillenniumDB & FalkorDB & Blazegraph \\
\hline
\hline
\rowcolor{Gray}
Total, ms & 24 403 & 166 882 & 34 891 & 276 527 & 460 606 & 403 738 \\
\hline
Mean, ms & 49.8 & 340.6 & 71.2 & 564.3 & 940.0 & 824.0 \\
\rowcolor{Gray}
Median, ms & 2.0 & 3.3 & 1.1 & 13.7 & 17.6 & 128.0 \\
\hline
Mean speedup & 1.00 & 0.15 & 0.70 & 0.09 & 0.05 & 0.06 \\
\rowcolor{Gray}
Median speedup & 1.00 & 0.60 & 1.74 & 0.15 & 0.11 & 0.02 \\
\hline\hline
Total C, ms & 35 061 & 508 378 & 288 766 & 639 193 & --- & --- \\
\hline
\rowcolor{Gray}
Mean C, ms & 427.6 & 6 199.7 & 3 521.5 & 7 795.0 & --- & --- \\
Median C, ms & 169.1 & 372.9 & 142.9 & 3 105.6 & --- & --- \\
\hline
\rowcolor{Gray}
Mean speedup C & 1.00 & 0.07 & 0.12 & 0.05 & --- & --- \\
Median speedup C & 1.00 & 0.45 & 1.18 & 0.05 & --- & --- \\
\hline\hline
\rowcolor{Gray}
Memory, Gb & 9.2 & 6.9 & 9.2 & --- & --- & --- \\ 
BPT & 16.3 & 12.1 & 16.3 & --- & --- & --- \\
\hline
\end{tabular}    
\end{table*}

\begin{table}
    \centering
    \caption{Yago-2S query execution results. Mean and median speedup relatively the proposed solution is a ratio of mean and median over query set for respective systems}
    \label{yago-results-table}
\resizebox{8.6cm}{!}{
\begin{tabular}{|l|c|c|c|c|c|}
\hline
 & {\footnotesize LARPQ} & {\footnotesize RPQ-matrix} & \CellWithForceBreak{{\scriptsize RPQ-matrix} \\ {\footnotesize (GrB)}} & {\footnotesize MillenniumDB}  & {\footnotesize Blazegraph} \\
\hline\hline
\rowcolor{Gray}
Total, ms & 377 & 174 & 202 & 7 025 & 24883 \\
\hline
Mean, ms & 54 & 25 & 29 & 1 004 & 3555 \\
\rowcolor{Gray}
Median, ms & 69 & 25 & 38 & 985 & 3808 \\
\hline
Mean speedup & 1.00 & 2.17 & 1.87 & 0.05 & 0.02\\
\rowcolor{Gray}
Median speed up & 1.00 & 2.71 & 1.81 & 0.07 & 0.02\\
\hline\hline
Memory, Gb & 0.5 & 0.4 & 0.5 & --- & --- \\
\rowcolor{Gray}
BPT & 11.3 & 9.8 & 11.3 & --- & --- \\
\hline
\end{tabular}
}
\end{table}

We measure query evaluation time with a 1-minute timeout, preloading all required data into memory, so data preprocessing time is not included. Queries execute sequentially in isolation.

\noindent\textbf{RPQ-Matrix\footnote{RPQ-Matrix repository we use for the evaluation: \url{https://github.com/adriangbrandon/rpq-matrix/tree/34fc2240a7c8069f7d6a39f1c75176edac4fe606}} configuration.}
\begin{itemize}
    \item CSR/CSC mode as the most performant one.
    \item No CPU/cores limitations.
    \item No memory limitations.
    \item Timing: internal timer.
\end{itemize}

\noindent\textbf{RPQ-matrix (GrB)\footnote{RPQ-matrix implemented using SuiteSparse:GraphBLAS repository: \url{https://github.com/suvorovrain/rpq-matrix/tree/gbmod}} configuration.}
\begin{itemize}
    \item Pre-caching CSR/CSC matrices.
    \item No CPU/cores limitations.
    \item No memory limitations.
    \item Timing: internal timer.
\end{itemize}

\noindent\textbf{MillenniumDB\footnote{MillenniumDB repository we use for the evaluation: \url{https://github.com/MillenniumDB/MillenniumDB/tree/5190c0d9b07ca681328495b69c715af792513775}}  configuration.}
\begin{itemize}    
    \item SPARQL mode.
    \item Maximum CPU cores.
    \item Two execution runs (first warm-up excluded).
    \item Timing: internal timer for query optimization  + execution (parser excluded).
\end{itemize}

\noindent\textbf{FalkorDB (v4.2.0)\footnote{FalkorDB v4.2.0: \url{https://github.com/FalkorDB/FalkorDB/tree/v4.2.0}} configuration.}
\begin{itemize}
    \item Accessed via Python wrapper.
    \item No CPU/cores limitations.
    \item No query caching.
    \item Vertices index for efficient start/end vertex selection.
    \item Timing: internal database timer.
\end{itemize}

\noindent\textbf{Blazegraph (2.1.5)\footnote{Blazegraph v2.1.5: \url{https://github.com/blazegraph/database/tree/BLAZEGRAPH_RELEASE_2_1_5}} configuration.}
\begin{itemize}
    \item Accessed via Python wrapper.
    \item No CPU/cores limitations.
    \item No memory limitations.
    \item Timing: external timer.
    \item Queries are executed one by one with extra heatup to avoid out-of-memory.
\end{itemize}

All benchmarking scripts, configurations, and instructions are available in GitHub repository\footnote{Benchmark evaluation repository: \url{https://github.com/SparseLinearAlgebra/la-rpq}}.

\subsection{Real-World Graph Querying Results}

Resulting per-query evaluation time is provided in Figure~\ref{all-results} for queries that are supported on all of the competitors. Remaining complex queries ended up with a timeout on slower competitors or not expressible in Cypher are represented separately in Figure~\ref{complex-results}. The dotted lines represent means and the straight lines represent medians.
These numeric values of and total execution time are available in the table~\ref{all-results-table} for the relatively simple and complex queries (C) separately. For linear algebra-based competitors the size of the dataset loaded in-memory, and memory byte-per-triple (BPT) memory consumption values are provided.

For simple queries, LARPQ demonstrates the best mean with a speedup of $1.7 \times$ to $18.9 \times$. However, RPQ-matrix implemented with SuiteSparse:GraphBLAS demonstrates the best median time that is better $1.7 \times$ than that of LARPQ. It means that RPQ-matrix evaluates some of the queries faster whereas the BFS-based algorithm works better in general cases. Our hypothesis is that the proposed algorithm does not utilise the sparsity of some edge kinds enough. Both linear algebra-based approaches demonstrate better time in mean an in median in comparison to all of database management systems.

For complex queries, relations between different competitors are the same except that LARPQ mean is drastically lower than other competitors being $14.5\times$, $8.3\times$, $18.3\times$ less than those of the competitors. It means the proposed algorithm is capable of executing the most complex out of the complex algorithm faster than other approaches.




The results for the Yago-2S dataset are presented in table \ref{yago-results-table}. FalkorDB is excluded since every of the $7$ queries have resulted with a timeout.

Both RPQ-matrix algorithms demonstrate better time than the proposed LARPQ algorithm. It is likely due to the structure of the query. All of them use $a\;b\;c^+\;d^+$ patterns. Iterative structure of the BFS-based approach does not utilize efficient evaluation order that is important for such long queries of simple operations.

It is an interesting question for future research whether it is possible to combine ideas from two linear algebra based-solutions (proposed and RPQ-Matrix) to take the best from both of them.

\subsection{Synthetic Graph Querying Results}

We perform a detailed comparison of competitors for executing different query kinds by running a synthetic RPQBench. Its evaluation results are available in the table~\ref{rpqbench-all}. Each row represents distinct query kind. Total execution time of the randomly generated queries are presented in seconds for each competitor separately. Due to the lower overall performance demonstrated on the real-world datasets, MillenniumDB, FalkorDB, and Blazegraph are excluded from subsequent comparisons. We also provide edge statistics for each label kind in table~\ref{rpqbench-label-stats}.

\begin{table}
    \centering
    \caption{RPQBench dataset evaluation time of queries with randomly generated sources and destinations per each query kind in seconds}
  \label{rpqbench-all}
  \begin{tabular}{|l|l|r|r|r|}
  \hline
   & \CellWithForceBreak{{\footnotesize Query pattern,  single-source} \\ {\footnotesize (S) or single-destination (D)}} & {\footnotesize LARPQ} & {\footnotesize RPQ-matrix} & \CellWithForceBreak{{\footnotesize RPQ-matrix} \\ {\footnotesize (GrB)}}\\
  \hline\hline\rowcolor{Gray}
  1 & $a\;b\;c$, (S)  & 51 & 83 & 11 \\
  2 & $a\;b\;c$, (D)  & 32 & 11 & 12 \\
  \hline\hline\rowcolor{Gray}
  3 & $(a\;b\;c) \mid (c\;d\;d)$, (S)  & 59 & 89 & 24 \\
  4 & $(a\;b\;c) \mid (c\;d\;d)$, (D)  & 47 & 14 & 22 \\
  \hline\hline\rowcolor{Gray}
  5 & $d^*$, (S)  & 22 & 21 & 37 \\
  6 & $d^*$, (D)  & 19 & 19 & 23 \\
  \hline\hline\rowcolor{Gray}
  7 & $d^*\;e$, (S)  & 21 & 12 & 18 \\
  8 & $d^*\;e$, (D)  & 6 & 3 & 5 \\
  \hline\hline\rowcolor{Gray}
  9 & $d^+$, (S)  & 23 & 14 & 29 \\
  10 & $d^+$, (D)  & 18 & 14 & 20 \\
  \hline\hline\rowcolor{Gray}
  11 & $(a\;b)^*$, (S)  & 1 350 & 30 156 & 4 904 \\
  12 & $(a\;b)^*$, (D)  & 3 619 & 15 706 & 2 978 \\
  \hline\hline\rowcolor{Gray}
  13 & $f\;g\;(d \mid e)$, (S)  & 92 700 & 5 917 & 7 745 \\
  14 & $f\;g\;(d \mid e)$, (D)  & 2 193 & 301 & 486 \\
  \hline\hline\rowcolor{Gray}
  15 & $f\;g\;(d \mid e)^*$, (S)  & 149 644 & 2 870 709 & 2 053 090 \\
  16 & $f\;g\;(d \mid e)^*$, (D)  & 6 167 & 810 & 1 053 \\
  \hline\hline\rowcolor{Gray}
  17 & $(c \mid g)\;(d \mid e)$, (S)  & 36 & 3 & 7 \\
  18 & $(c \mid g)\;(d \mid e)$, (D)  & 16 & 173 017 & 41 713 \\
  \hline\hline\rowcolor{Gray}
  19 & $(c \mid g)\;(d \mid e)^*$, (S)  & 930 & 55 & 232 \\
  20 & $(c \mid g)\;(d \mid e)^*$, (D)  & 99 & 955 891 & 229 035 \\
  \hline
  \end{tabular}  
\end{table}

\begin{table}
    \centering
    \caption{RPQBench edge stats per label }
    \label{rpqbench-label-stats}
    \begin{tabular}{|l|r||l|r|}
    \hline
    Edge label & Count & Edge label & Count \\
    \hline\hline
    \rowcolor{Gray}
    $a$ & 343 660 & $e$ & 36\\
    $b$ & 4 209 447 & $f$ & 4 928 456\\ 
    \rowcolor{Gray}
    $c$ & 114 742 222 & $g$ & 223 656\\
    $d$ & 186 & & \\     
    \hline
    \end{tabular}    
\end{table}

As it is observed, the execution time of the proposed BFS-based algorithm for simple queries such as $d^*$, $d^+$, and $d^* e$ is quite similar to that of other linear algebra-based implementations. RPQ-matrix $\times 2$ speedups are likely to happen due to the CSR/CSC matrix implementation since RPQ-matrix (GrB) demonstrates evaluation time close to LARPQ.

The greatest performance improvement over other approaches is achieved when the patterns contain compound parts involving dense edges that are not adjacent to the starting or final node. For instance, the single-destination query $20$, $(c \mid g)\,(d \mid e)^*$, yields speedups of $9{,}600\times$ and $2{,}300\times$. The same holds for queries $11$, $15$, and $18$.

For the remaining queries, such as 1--4, the proposed algorithm demonstrates a relative slowdown of $2\times$ to $4\times$, since it does not utilize an efficient evaluation order. For queries 14, 16, 17, and 19, LARPQ is $4\times$ to $17\times$ slower than the competitors because it does not exploit the fact that edges with labels $d$ or $e$ are very rare.

Conclusion about various linear algebra-based algorithms might be summarized as follows.

\begin{itemize}
  \item LARPQ is a better choice for queries that involve compound operations over labels corresponding to many edges and not having source/destination vertices close to them.
  \item LARPQ algorithm tends to be more stable whereas RPQ-matrix might demonstrate drastic slowdown for some kinds of complex queries.
  \item Both algorithms are efficient enough for simple queries.
  \item RPQ-matrix is a better choice if the query contains rare labels, long concatenations or disjunctions allowing to perform optimizations.
\end{itemize}

\section{Conclusion and Future Work}

In this work, we proposed the single-source RPQ evaluation algorithm, which is based on the simultaneous traversal of the input graph and the finite automaton specifying path constraints.
The traversal is expressed in terms of operations over matrices and vectors, which allows us to provide a highly parallel implementation based on SuiteSparse:GraphBLAS. 

Our experimental evaluation shows that the suggested algorithm is suitable for in-memory processing of real-world large knowledge graphs. 
While for some queries the proposed algorithm is slower than competitors, we can conclude that our solution is faster for hard queries: it fits with 1 minute time while other solutions do not. 

Our results demonstrate that both algorithmic design and implementation details of underlying linear algebra primitives significantly impact 2-RPQ performance. While average performance metrics provide straightforward comparisons, specific cases require deeper analysis to identify each algorithm's strengths under different conditions; to understand implementation trade-offs; to guide development of robust universal solutions.

Furthermore, this analysis provides valuable insights for enhancing the GraphBLAS API by identifying critical functionality required for efficient RPQ evaluation algorithms. The performance characteristics we observed highlight specific linear algebra primitives that most significantly impact query processing efficiency, suggesting potential directions for API optimization.

Regarding our new algorithm, first of all, it is necessary to analyse abilities to apply well-known optimizations from both RPQ evaluation and BFS algorithms. 
For example, rare labels~\cite{10.1007/978-3-642-31235-9_12} utilization, or push-pull optimization~\cite{10.1145/3466795} respectively.

Although multiple source BFS has been shown to be expressed in terms of linear algebra~\cite{9286186}, there is room for technical optimizations and careful evaluation of the respective modifications to the proposed algorithm, and it should be done in the future.

Thanks to linear algebra, having a single schema of algorithm one can solve different problems varying semiring-like structures. For example, one can look at variations of the BFS~\cite{brock2021graphblas} where one can compute reachability facts or information about paths depending on the used semiring. In the case of RPQ, there are a number of possible \emph{outputs} and \emph{semantics}~\cite{10.1145/3104031}: reachability, single path, all paths, simple paths, etc. It is an open question, which of them can be expressed without algorithm changes, but by providing other semirings, and which can be expressed with algorithm changes.

Utilization of GPGPUs to evaluate linear algebra-based algorithms for graph analysis can significantly improve performance in some cases~\cite{10.1145/3466795, stanislavovic2023generalized}.
It is necessary to investigate, whether utilization of GPGPU in our algorithm improves performance or not.

Distributed solutions are a way to process graph processing~\cite{10.1007/s10115-020-01536-2}. 
Implementation and evaluation of the proposed algorithm in distributed settings, for example, using CombBLAS~\cite{doi:10.1177/1094342011403516} that provides distributed linear algebraic routines for graph analysis, is also a task for the future.

\begin{acks}
This research has been supported by the St. Petersburg State University, grant id 116636233. Paper submission has been sponsored by Tarantool DBMS team, VK Tech LLC.
\end{acks}


\bibliographystyle{ACM-Reference-Format}
\bibliography{ss-la-rpq}

\appendix

\section{Proof of Correctness}\label{appendix:proof-of-correctness}

\begin{theorem}[LA 2-RPQ algorithm correctness] 
The proposed algorithm, represented in~\ref{ssr-algorithm}, computes the matrix $P$ such that the respective relation $\mathcal{P} \subseteq Q \times V$ has the following property.
\[
(q, v) \in \mathcal{P} \Leftrightarrow 
\begin{cases}
    \exists \text{ 2-path } \pi_G \text{ in } \mathcal{ G } \text{ from } v_s \text{ to } v \\
    \exists \text{ path } \pi_N \text{ in } \mathcal{N} \text{ from some } q_s \in Q_S \text{ to } q \\
    \omega^\leftrightarrow_\mathcal{G}(\pi_\mathcal{G}) \cap \omega_\mathcal{N}(\pi_\mathcal{N}) \ne \oemptyset
    \tag{$\ast$}\label{P-inv}.
\end{cases}
\]
\end{theorem}

Denote the state of matrix $M$ after the step $n$ of the loop on lines 7--10 of algorithm~\ref{ssr-algorithm} as $M_n$ and the state of $P$ as $P_n$. Introduce the auxiliary relations $\mathcal{M}_n \subseteq Q \times V$ represented by $M_n$ and $\mathcal{P}_n \subseteq Q \times V$ represented by $P$ after step $n$. $\mathcal{P}$ and $\mathcal{P}_n$ are connected with relations $\mathcal{M}_n$:

\[
\mathcal{P}_n = \bigcup_{1 \le m \le n} \mathcal{M}_m;\;\; \mathcal{P} = \bigcup_{m \in \mathbb{N}} \mathcal{M}_m
\]

The approach is to build a relation between the automaton states and the vertices of the graph and update it by traversing both automaton and graph at the same time. As an algorithm invariant after the step $n$ we claim this properties for $\mathcal{M}_n$ represented by the matrix $M_n$:
\[
(q, v) \in \mathcal{M}_n \Leftrightarrow
\begin{cases}
    \exists \text{ 2-path } \pi_\mathcal{G} \text{ of length } n \text{ in } \mathcal{ G } \text{ from } v_s \text{ to } v \\
    \exists \text{ path } \pi_\mathcal{N} \text{ of length } n \text{ in } \mathcal{N} \text{ from } q_s \in Q_S \text{ to } q \\
    \omega^\leftrightarrow_\mathcal{G}(\pi_\mathcal{G}) \cap \omega_\mathcal{N}(\pi_\mathcal{N}) \ne \oemptyset \\
    \forall m < n\;(q, v) \notin \mathcal{M}_m
    \tag{$\ast\ast$}\label{M-inv}
\end{cases}
\]

Note that if we continue looping forever in lines 7--10 from $\mathcal{M}_n = \oemptyset$ it follows $\mathcal{M}_m = \oemptyset$ for $m > n$. This means that it is enough to iterate until the $M$ matrix becomes $\mzero$. 

Notice that $\mathcal{P}_{n} \subsetneq \mathcal{P}_{n+1}$ or $|\mathcal{P}_n| < |\mathcal{P}_{n+1}|$ if $\mathcal{M}_{n+1}$ is not empty. And $\mathcal{M}_{n+1} \cap \mathcal{P}_n = \oemptyset$ for all $n \in \mathbb{N}$. $|\mathcal{P}_n| \leq |Q||V|$ and $|\mathcal{M}_n| \leq |Q||V|$ for all $n \in \mathbb{N}$. Thus, if $\mathcal{M}_m \ne \oemptyset$ for all $m < |Q||V|$ then $\mathcal{M}_{|Q||V|} = \oemptyset$, since $|\mathcal{P}_{|Q||V| - 1}| \geq |Q||V|$. This means that the algorithm always finishes in a maximum of $|Q||V|$ steps.

Obviously, the invariant holds for $n = 0$ after initializing the matrix $M$ with $M_0$. If you consider the paths of length $0$, the set of vertices and automaton states coincides with $\{ v_s \}$ and $Q_S$ correspondingly.

\[
(q, v) \in \mathcal{M}_0 \Leftrightarrow q \in Q_S, v = v_s
\]

Consider evaluating the $n+1$ step of the algorithm. Fix the label $a \in L$. After step $n$, $M$ represents a relation $\mathcal{M}_n$. At first, evaluate the first matrix product $M'^a_{n+1} = (N^a)^T \mtimes M$. This product represents a relation $\mathcal{M}'^{a}_{n+1} \subseteq Q \times V$:
\[
 \mathcal{M}'^{a}_{n+1} = \{ (q', v) \mid (q, q') \in \Delta^a, (q, v) \in \mathcal{M}_n \}.
\]
Evaluate the second matrix product $M^a_{n+1} = M'^{a}_{n+1} \mtimes G^a =$\\$= (N^a)^T \mtimes M_n \mtimes G^a$. Assume $M^a_{n+1}$ represents a relation $\mathcal{M}^a_{n+1}$, then:
\begin{align*}
\mathcal{M}_{n+1}^a &= \{ (q', v') \mid (v, v') \in E^a, (q',v) \in \mathcal{M}'^{a}_{n+1} \} \\
&= \{ (q', v') \mid (v, v') \in E^a, (q, q') \in \Delta^a, (q, v) \in \mathcal{M}_n \}. 
\end{align*}

Hence, the new relation $\mathcal{M}_{n+1}$ represented by the matrix \\ $ \bigvee_{a \in \Sigma^\leftrightarrow \cap L^\leftrightarrow} M^a_{n+1} \langle \neg P \rangle$ can be written as follows:

\begin{align*}
\mathcal{M}_{n+1} & = \bigcup_{a \in \Sigma^\leftrightarrow \cap L^\leftrightarrow} \{ (q, v) \mid (q, v) \in \mathcal{M}'^{a}_{n+1}, (q, v) \notin \mathcal{P} \} = \\
&= \bigcup_{a \in \Sigma^\leftrightarrow \cap L^\leftrightarrow} \left\{ (q', v') \middle| 
\begin{gathered}
(v, v') \in E^a, (q, q') \in \Delta^a \\
(q, v) \in \mathcal{M}_n, (q, v) \notin \mathcal{P} 
\end{gathered}
\right\}.
\end{align*}

Updating the relation between the 2-NFA states and the graph vertices $\mathcal{M}$ with the $\mathcal{M}'$ value derives the following properties.
\[
(q', v') \in \mathcal{M}_{n+1} \Leftrightarrow 
\begin{cases}
(q, q') \in \Delta^a \\ 
(v, v') \in E^a \\
\end{cases}
\text{ for some } a \in \Sigma^\leftrightarrow \cap L^\leftrightarrow.
\]

Ensure that the invariant is preserved for $(q', v')$ in $\mathcal{M}'$. Let $\pi_\mathcal{G} = (e_1, ..., e_n)$, $\pi_\mathcal{N} = (\delta_1, ..., \delta_n)$ be the paths to $(q, v)$ satisfying conditions \ref{M-inv}. Then:
\begin{itemize}
    \item $\exists$ 2-path $\pi_\mathcal{G}' = (e_1, ..., e_n, (v, v'))$ in $\mathcal{G}$ from $v_s$ to $v'$.
    \item $\exists$ path $\pi_\mathcal{N}' = (\delta_1, ..., \delta_n, (q, q'))$ in $\mathcal{N}$ from $q_F \in Q_F$ to $q'$.
    \item $w \cdot a \in \omega^\leftrightarrow_\mathcal{G}(\pi_\mathcal{G}') \cap \omega_\mathcal{N}(\pi_\mathcal{N}')$ where $w \in \omega^\leftrightarrow_\mathcal{G}(\pi_\mathcal{G}) \cap \omega_\mathcal{N}(\pi_\mathcal{N})$.
    \item $\mathcal{P}_n = \displaystyle \bigcup_{m \le n} \mathcal{M}_m$, so if $(q, v) \in \mathcal{M}_m $ for $m \le n$ then $(q, v) \notin \mathcal{M}_{n+1}$.
\end{itemize}

Since conditions \eqref{M-inv} are preserved for the relation $\mathcal{M}_{n+1}$ matching the new value of $M \gets M_{n+1}$, the invariant is preserved.

\balance 

\end{document}